\documentclass[twocolumn,showpacs,amsmath,amssymb,pra,floatfix]{revtex4}
\usepackage{graphicx}% Include figure files
\usepackage{dcolumn}% Align table columns on decimal point
\usepackage{bm}% bold math
\usepackage{bbm}
\usepackage{multirow,amssymb,amsbsy,amsmath}
\usepackage{stmaryrd}
\newcommand{\ddim}{\udelta\kern0.1em}

\newcommand{\beikonst}[2]{\left( #1 \right)_{\kern-0.2em #2}}
\newcommand{\tr}[2][]{\text{Tr}_{#1}\left\{#2\right\}}

\newcommand{\trtxt}[2][]{\text{Tr}_{#1}\{#2\}}

\begin{document}
%\linenumbers

%\preprint{APS/123-QED}

% -----------------------------------------
%
% Title
%

\title{Variational analysis of driven-dissipative Rydberg gases}

\author{Hendrik Weimer}%
\email{hweimer@itp.uni-hannover.de}
\affiliation{Institut f\"ur Theoretische Physik, Leibniz Universit\"at Hannover, Appelstr. 2, 30167 Hannover, Germany}

\date{\today}% 

\begin{abstract}

  We study the non-equilibrium steady state arising from the interplay
  between coherent and dissipative dynamics in strongly interacting
  Rydberg gases using a recently introduced variational method
  [H. Weimer, Phys. Rev. Lett. \textbf{114}, 040402 (2015)]. We give a
  detailed discussion of the properties of this novel approach, and we
  provide a comparison with methods related to the
  Bogoliubov-Born-Green-Kirkwood-Yvon hierarchy. We find that the
  variational approach offers some intrinsic advantages, and we also
  show that it is able to explain the experimental results obtained in
  an ultracold Rydberg gas on an unprecedented quantitative level.

\end{abstract}

% PACS, the Physics and Astronomy Classification Scheme.

\pacs{32.80.Ee, 03.65.Yz, 05.30.Rt }
\maketitle

\section{Introduction}

The discovery of dissipative quantum state engineering
\cite{Diehl2008,Verstraete2009,Weimer2010}, i.e., the use of
controlled sources of dissipation for the preparation of many-particle
quantum states, has led to a surge of interest in open quantum
many-body systems. The unrivaled tunability of interaction and
dissipation properties of driven Rydberg gases makes them particularly
useful for this purpose and has resulted in a large amount of
theoretical and experimental works investigating the interplay between
coherent and dissipative dynamics in these systems
\cite{Raitzsch2009,Lee2011,Honer2011,Weimer2011,Glatzle2012,Ates2012a,Lemeshko2013a,Carr2013,Rao2013,Carr2013a,Hu2013,Honing2013,Otterbach2014,Sanders2014,Schonleber2014,Malossi2014,Hoening2014,Marcuzzi2014,Qian2014,Urvoy2014,Weber2015}. Here,
we provide a detailed discussion of the application of a recently
introduced variational principle for the steady state of dissipative
quantum many-body systems \cite{Weimer2015} to driven-dissipative
Rydberg gases.

Rydberg atoms are routinely excited from ground state atoms by
coherent laser driving, with their atomic properties and interaction
strength scaling dramatically with the principal quantum number
\cite{Low2012}. The radiative decay of the metastable Rydberg state
provides a natural dissipative element whose decay rate can be widely
tuned by laser coupling to other excited non-Rydberg states
\cite{Zhao2012}. As such, Rydberg atoms provide an ideal environment
for studying dissipative many-body dynamics in a strongly interacting
regime.

In this article, we perform an analysis of the non-equilibrium steady
state of a driven-dissipative Rydberg gas. We investigate the
properties of this steady state using a variational method recently
introduced by the author \cite{Weimer2015}. We provide a detailed
evaluation of the approximations carried out within this novel
method. We complement our analysis of the variational method by
describing an alternative approach to the investigation of dissipative
quantum many-body systems based on a hierarchy of equation that also
allows for a systematic incorporation of correlations. We perform an
explicit comparison of the two methods, finding substantial advantages
in favor of the variational approach. Finally, we demonstrate that the
variational method provides remarkable quantitative agreement with the
experimental results on the steady state phase transition observed in
an ultracold Rydberg gas.

\section{Driven-dissipative Rydberg gases}

We first give a microscopic description of driven-dissipative Rydberg
gases. We express the dynamics in terms of a spin $1/2$
representation, where the down spin state corresponds to an electronic
ground state, and the up spin state refers to a highly excited Rydberg
state \cite{Low2012}. As the frequency difference between all
electronic states is much larger than the decay rate of the Rydberg
state, it is well justified to describe the radiative decay of the
Rydberg excitations in terms of a Markovian quantum master equation in
Lindblad form,
\begin{equation}
  \frac{d}{dt}\rho = -\frac{i}{\hbar}[H,\rho] + \sum\limits_{i} \left(c_i\rho c_i^\dagger - \frac{1}{2}\left\{c_i^\dagger c_i, \rho\right\}\right),
\end{equation}
where $\rho$ is the density operator describing the state of the
system, $H$ is the Hamiltonian accounting for the coherent part of the
dynamics, and the set of $c_i$ are quantum jump operators responsible
for the dissipation \cite{Breuer2002}. If the external laser fields
are close to a resonance between the atomic ground state and a single
Rydberg state, such a system in its rotating frame is governed by the
spin $1/2$ Hamiltonian
\begin{equation}
  H = -\frac{\hbar\Delta}{2} \sum\limits_i \sigma_i^z + \frac{\hbar\Omega}{2}\sum\limits_i\sigma_i^x + \sum_{i<j} \frac{C_6}{|{\bf r}_i - {\bf r}_j|^6} P^{r}_i P^{r}_j,
\end{equation}
where the Rabi frequency $\Omega$ and the detuning $\Delta$ represent
the laser parameters, the $C_6$ coefficient denotes the strength of
the van der Waals repulsion between Rydberg states, and $P^r_i =
(1+\sigma^z_i)/2$ is the projection onto the Rydberg state. The jump
operators describing the decay of the Rydberg excitations is
represented by quantum jump operators describing up spins flipping
into down spins according to $c_i = \sqrt{\gamma}\sigma_-^{(i)}$, with
$\gamma$ being the decay rate of the Rydberg state. Initially, we will
assume that the system is well described by a model involving only
nearest-neighbor interactions, i.e., the blockade radius is smaller
than the lattice spacing $a$. Then, we can introduce an interaction
constant given by $V = C_6/a^6$. Of special interest is the regime
where the Hamiltonian is equivalent to a transverse field Ising model,
which is realized for $\Delta = V/2$.

In contrast to closed quantum systems, an open quantum systems such as
the dissipative Rydberg gas will generically relax towards a
stationary state characterized by the condition $\frac{d}{dt} \rho =
0$. Crucially, the interplay between coherent and incoherent part of
the dynamics generically leads to a stationary state that is different
from any state in thermal equilibrium, i.e., a non-equilibrium steady
state.

\section{Variational principle}

In the following, we review the basic concepts behind the variational
principle introduced in Ref.~\cite{Weimer2015}. The basic idea is to
take a variational trial state $\rho$ and compute the residual
dynamics it will generate by computing its time-derivative
$\dot{\rho}=\mathcal{L}\rho$ according to the underlying quantum
master equation with the Liouvillian superoperator $\mathcal{L}$. The
operator $\dot{\rho}$ can be expressed as a traceless Hermitian
matrix. To find the variational approximation to the true steady state
having $\dot{\rho}=0$, we choose the variational state that minimizes
the trace norm of $\dot{\rho}$, i.e.,
\begin{equation}
  \rho_{var} = \operatorname*{\arg\,\min}_\rho \trtxt{|\mathcal{L}\rho|}.
  \label{eq:principle}
\end{equation}
Choosing the trace norm as the correct matrix norm can be motivated on
two different grounds. First, the trace norm is unbiased in the sense
that it does not favor certain classes of variational states over
others without a physical reason. This is related to the linearity
condition $||\dot{\rho}||=||\lambda \dot{\rho}||/\lambda$ satisfied by
the trace norm. In particular, any other Schatten norm $||\dot{\rho}||
= \trtxt{|\dot{\rho}|^p}$ with $p>1$ is biased towards the maximally
mixed state \cite{Weimer2015}.

The second way to motivate the choice of the trace norm follows
follows from quantum information theory. Here, the trace norm can be
interpreted as being equivalent to the trace distance of $\dot{\rho}$
to the zero matrix, while the latter is obtained for $\dot{\rho}$ if
and only if the variational state $\rho$ is an exact stationary state
of the master equation. Importantly, the ability to physically
distinguish the operator $\dot{\rho}$ from the zero matrix is given by
their trace distance \cite{Gilchrist2005}, and hence the trace norm of
$\dot{\rho}$. In this sense, the trace norm is the natural norm to
decide which variational state is the best approximation to the true
steady state.  We would also like to point out that the trace norm for
the steady state is equivalent to applying a time-dependent
variational principle for the dynamical evolution \cite{Kraus2012}.

Following these initial statements, we can now proceed with the
variational analysis. Calculating the trace norm is in general still a
computational problem scaling exponentially with the system size, so
additional steps are needed first. This situation is very similar to
that of correlated fermions, where energy expectation values for
variational Gutzwiller wave functions can only be evaluated within
further approximations \cite{Edegger2007}. In our case, we exploit the
fact that we are not so much interested in the actual value of the
trace norm, but rather in the properties of the stationary state. The
additional approximations we will carry out retain the variational
character of our calculation, i.e., they provide a rigorous upper
bound to the trace norm.

\subsection{Product states}

To be explicit, we first consider the case of the variational set of states
being restricted to product states, i.e.,
\begin{equation}
    \rho = \mathcal{R} 1 = \prod\limits_i \rho_i.
\end{equation}
Here, we have introduced the superoperator $\mathcal{R}$, which
replaces every occurrence of the identity operator for site $i$,
$1_i$, by the single-site density matrix $\rho_i$. Additionally, we
focus on quantum master equations including nearest-neighbor
interactions or jump operators involving at most two adjacent
sites. Then, the trace norm of the resulting dynamics can be written
in the form
\begin{equation}
  ||  \dot{\rho}|| = ||\sum\limits_i \mathcal{R}\dot{\rho}_i +\sum\limits_{\langle ij \rangle} \mathcal{R} \dot{C}_{ij}||,
\end{equation}
where $\dot{\rho}_i = \trtxt[\not i]{\dot{\rho}}$ describes the
single-site dynamics and $C_{ij}$ accounts for correlations between
the sites. Here, the correlations between the sites stem from the
nearest-neighbor interactions and two-site jump operators and thus are
restricted to nearest-neighbor correlations only. As a first
approximation, we apply the triangle inequality to pull the summation
over $i$ out of the norm,
\begin{equation}
  ||  \dot{\rho}|| \leq \sum\limits_i ||\mathcal{R}\dot{\rho}_i + \sum\limits_j \mathcal{R}\dot{C}_{ij}||.
\end{equation}
As the next step, we make use of the fact that $\dot{\rho}_i$ and
$\dot{\rho}_j$ act on different parts of the Hilbert space, which
allows us to write
\begin{equation}
  ||  \dot{\rho}|| \leq \sum\limits_i ||\mathcal{R}\dot{\rho}_i + \sum\limits_j \mathcal{R}\left(\rho_i\dot{\rho}_j + \dot{C}_{ij}\right)||.
\end{equation}
Note that this inequality does not change the variational
approximation to the steady state, but it allows us to write the final
result in a more compact form. Finally, we employ the triangle
inequality a second time, yielding
\begin{equation}
\label{eq:ineq}
  ||  \dot{\rho}|| \leq \sum\limits_{\langle ij\rangle} ||\mathcal{R}\left(\dot{\rho}_i\rho_j + \rho_i\dot{\rho}_j + \dot{C}_{ij}\right)|| =  \sum\limits_{\langle ij\rangle} \trtxt{|\dot{\rho}_{ij}|} 
\end{equation}
Consequently, we have succeeded in mapping the full quantum many-body
problem into a sum of efficiently solvable problems involving only
neighboring sites. For translationally invariant systems it is in
general even sufficient to solve a single two-site problem.

\begin{figure}[t]

\includegraphics[width=\linewidth]{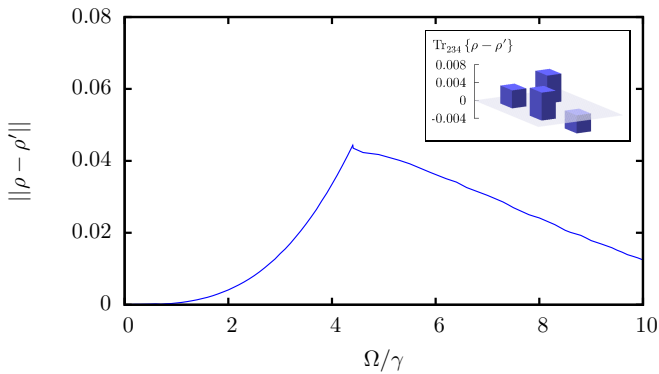}

\caption{Difference in the variational states introduced by the
  inequalities leading to Eq. (\ref{eq:ineq}) as indicated by the trace
  distance between the states $\rho$ and $\rho'$. Here the full space
  density matrix $\rho$ was taken as a product state of four
  sites. The inset shows the difference in the single particle reduced
  density matrices at $\Omega=4\gamma$ ($\Delta = V/2$, $V=5\,\gamma$).}

\label{fig:ineq}

\end{figure}

At this point, it becomes a natural question to ask how well justified
our approximations are. Besides the obvious approximation in
restricting to the variational manifold, we have to assess the
deviations introduced by the inequalities leading to
Eq.~(\ref{eq:ineq}). For small system sizes, we can answer this
question exactly because the Hilbert space is still small enough so
that we can minimize the trace norm of Eq.~(\ref{eq:principle})
without any additional approximations. As shown in
Fig.~\ref{fig:ineq}, the deviation introduced by the inequalities is
quite small for the Ising model describing the dissipative Rydberg gas
and in the case of four particles.

\begin{figure}[t]

\includegraphics[width=\linewidth]{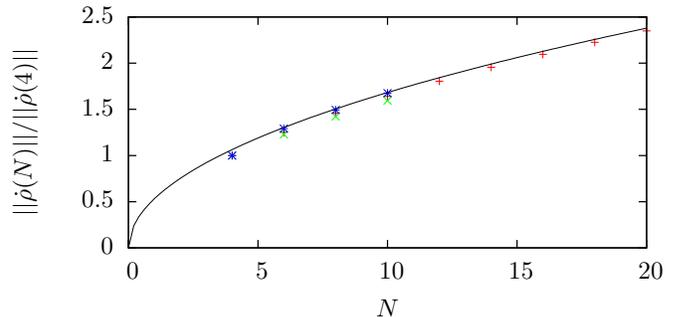}

\caption{Trace norm of the master equation depending on the system
  size $N$ for different models and different product states
  normalized to the value at $N=4$. The asymptotic behavior for the
  noninteracting decay model is shown as a solid line.}

\label{fig:scaling}

\end{figure}

Of course, we are not really interested in problems involving four
sites, but rather see how the situation behaves in the thermodynamic
limit. For this, we can check how the exact variational norm behaves
as a function of system size. As a single evaluation of the norm is
computationally less costly than performing a full minimization, we
can go to somewhat larger system sizes. Remarkably, we find that the
scaling with system size is completely independent of the model being
investigated or the trial state for which the norm is evaluated, see
Fig.~\ref{fig:scaling}. For a simple non-interacting toy model, this
surprising fact can be understood as a consequence of the central
limit theorem. This model consists of $N$ purely dissipative two-level
systems, whose jump operators are given by a dissipative spin-flip of
the form $c_i = \sqrt{\gamma}\sigma_i^-$, as in the case of the
dissipative Rydberg gas. As a trial state, we choose the maximally
mixed state, $\rho = 1/2^N$. Since the model is purely classical and
non-interacting, we can give an analytical expression for the trace
norm of the master equation,
\begin{equation}
  ||\dot{\rho}|| = 4\,\sum\limits_{m=0}^{N/2}\left(\begin{array}{cc}N\\\frac{N}{2}-m\end{array}\right)\frac{m}{2^N}\,\gamma.
\end{equation}
This expression can be evaluated efficiently even for large values of
$N$ and corresponds to the crosses in Fig.~\ref{fig:scaling}. To
obtain the asymptotic behavior in the limit of large $N$, we replace
the sum by an integral and use the central limit theorem to
approximate the binomial coefficients by a Gaussian function,
obtaining
\begin{align}
  ||\dot{\rho}|| &\approx \int\limits_0^{N/2} \frac{4}{\sqrt{\pi N/2}}\exp\left[-\frac{\left(x-\frac{N}{2}\right)^2}{\frac{N}{2}}\right]\,\left(\frac{N}{2}-x\right)\,\gamma\,dx\nonumber\\
  &= \sqrt{\frac{2N}{\pi}}\gamma + O(\sqrt{N\exp[-N]}).
\end{align}
As shown in Fig.~\ref{fig:scaling}, this asymptotic behavior is
already reached for quite small values of $N$, indicating that the
central limit theorem can also be applied to the fully quantum case,
which appears to be a natural consequence of the correlations in
$\dot{\rho}$ being restricted to nearest neighbors. Therefore, we
conclude that the approximations needed for an efficient calculation
of the variational norm are well justified and enable to use the
variational principle as a powerful tool to compute steady state
properties.

\subsection{Correlations}

The considerations made for product states can also be extended
towards more generic classes of variational states including
correlations. Here, we study the case where nearest-neighbor
correlations are fully included, resulting in a variational state
according to
\begin{equation}
  \rho = \prod_i \rho_i + \sum\limits_{\langle ij \rangle} \mathcal{R} C_{ij} + \sum\limits_{\langle ij\rangle \ne \langle kl\rangle}\mathcal{R}C_{ij}C_{kl}+\ldots.
\end{equation}
Additionally, we impose the constraint that all reduced density
matrices $\rho_{ij} = \rho_i\rho_j + C_{ij}$ are positive
definite. Using the same steps as for product states and observing
that partial traces of correlations vanish, $\trtxt[i]{C_{ij}}=0$, we
can again find an upper bound to the variational norm as
\begin{equation}
||\dot{\rho}|| \leq \sum\limits_{\langle ijk \rangle} || \dot{\rho}_{ijk}||.
\end{equation}
Here, the many-body problem reduces to a sum of three-site
problems. It should not be surprising that the inclusion of all
two-site correlations leads to the minimization of a three-site problem
as the interaction generically generates one higher order of
correlations. Consequently, treating $n$-body correlations exactly
requires to solve an $n+1$-body minimization problem.

\section{Hierarchy equation methods}

An alternative way to analyze dissipative many-body dynamics is
through a hierarchy of equations in terms of their correlations, in
close analogy to the Bogoliubov-Born-Green-Kirkwood-Yvon hierarchy of
classical physics \cite{Liboff2003}. For this, we express the density
operator $\rho$ in terms of their reduced density matrices, according
to the generating functional
\begin{equation}
  \mathcal{F}(\alpha) = \log \tr{\rho \prod\limits_i (1_i + \alpha_i)},
\end{equation}
where the operators $\alpha_i$ form an arbitrary operator basis acting
on lattice site $i$ \cite{Navez2010}. The first terms of the hierarchy
are then given by
\begin{align}
  \rho_{i} &= \tr[\not i]{\rho} = \left.\frac{\partial\mathcal{F}}{\partial \alpha_i}\right|_{\alpha=0}\\
  \rho_{ij} &= \tr[\not i\not j]{\rho} = \rho_i\rho_j + \underbrace{\left.\frac{\partial^2\mathcal{F}}{\partial \alpha_i\partial \alpha_j}\right|_{\alpha=0}}_{C_{ij}}\\
  \rho_{ijk} & = \tr[\not i\not j\not k]{\rho} = \rho_i\rho_j\rho_k + C_{ij}\rho_k + C_{ik}\rho_j + C_{jk}\rho_i\nonumber\\& + \left.\frac{\partial^3\mathcal{F}}{\partial \alpha_i\partial \alpha_j\partial\alpha_k}\right|_{\alpha=0}.
\end{align}
The same analysis can also be performed on the level of the quantum
master equation to obtain effective equations of motion for the
reduced density matrices. Generically, each equation of motion is
coupled to the next higher equation of motion in the hierarchy. The
usual strategy is then to truncate the hierarchy at some point by
setting the contributions from all higher order derivatives of
$\mathcal{F}$ to zero and solve the resulting closed set of equations
\cite{Navez2010}. This approximation is attributed to a $1/z$
suppression of the higher order derivatives. In lowest order, the
resulting equations of motion are identical to the mean-field
decoupling of Ref.~\cite{Diehl2010}.

In principle, it is possible to systematically incorporate
correlations similar as in the variational approach by going up to
higher terms in the hierarchy. However, the main drawback of the
method remains, that it cannot be formulated in terms of a variational
principle, i.e., the neglected higher order terms are uncontrolled. In
the case of the hierarchy equations having multiple solutions, it is
actually possible to combine it with the variational method. First,
all solutions to the hierarchy equations are computed, which are then
used as a variational class of states to find the solution that leads
to a minimization of the variational norm.

\section{Results}

\subsection{Lattice model}

\label{sec:lat}

\begin{figure*}[ht]

  \includegraphics{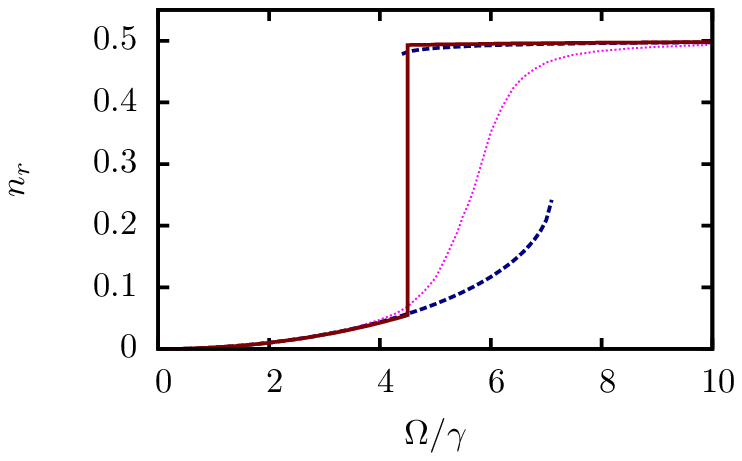} \includegraphics{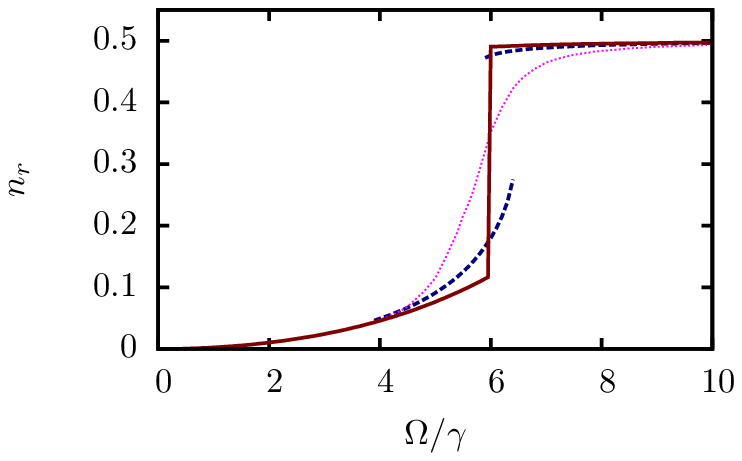}

  \caption{Comparison of the Rydberg density $n_r$ for the variational
    solution (solid) and the solutions of the hierarchy equations
    (dashed) for product states (left) and correlated states
    (right). For reference, the quantum trajectories solution of the
    quantum master equation for a $4\times 4$ system is shown as a
    dotted line. The data for both the variational and quantum
    trajectories solutions is taken from Ref.~\cite{Weimer2015}
    ($\Delta = V/2$, $V=5\,\gamma$).}

\label{fig:comp}

\end{figure*}

We will first put our attention to the case where the atoms are
distributed on a two-dimensional square lattice. We will make a direct
comparison between the variational method and the results from solving
the hierarchy equations. We complement this comparison with results
from a numerical solution of the quantum master equation using a
quantum trajectories method \cite{Johansson2013}. To ensure a
comparison on an equal footing, we will compare the variational
results for product states to the first order hierarchy equations, and
the variational method for correlated states to the second order
hierarchy equations. In the latter case, we include only
nearest-neighbor correlations within both methods. The results are
shown in Fig.~\ref{fig:comp}. Except for intermediate values of
$\Omega$, the two methods agree very well, and for correlated states,
the level of agreement is further improved and also matches well with
the results from the quantum trajectories simulation.

However, some important qualitative differences remain even when
correlations are included. The bistability of the first order
hierarchy solution is still present, although over a smaller range of
parameters. Consequently, this bistability is not just an artifact of
the first order result and the approximation of neglecting the $1/z$
corrections to it. Rather, it appears to be a generic element of the
hierarchy equation method. In contrast, the variational solution
always produces a unique stationary state, as even in the case of
multiple local minima of the variational norm, there is always a
unique global minimum. Within the variational approach, we find a
first-order phase transition between a low-density gas of Rydberg
excitations and a high-density liquid \cite{Weimer2015}.

\begin{figure}[b]

  \includegraphics[width=\linewidth]{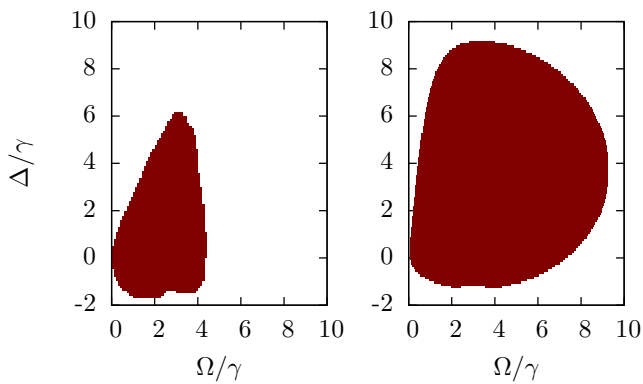}

  \caption{Extension of the antiferromagnetic phase. The shaded area
    depicts the presence of antiferromagnetic order according to the
    variational approach (left) and the first-order hierarchy
    equations (right). ($V=5\,\gamma$) }

\label{fig:af}

\end{figure}

Finally, we turn to the parameter regime for nonzero detuning $\Delta$
where the mean-field solution (i.e, the first-order hierarchy
equation) predicts the existence of an antiferromagnetic phase
\cite{Lee2011}. Within the variational method, we also find such an
antiferromagnetic phase, see Fig.~\ref{fig:af}, but its extension is
reduced significantly.

\subsection{Infinite dimension limit}

The qualitative differences between the variational approach and the
hierarchy equation method warrants further discussion, especially
regarding the appearance of the bistable region. Some previous works
have interpreted this region as a genuine thermodynamic phase
\cite{Lee2011,Marcuzzi2014}. In this context, the concept of the
\emph{lower critical dimension} is particularly important. It refers
to the spatial dimension of a system above which a phase transition
can be observed \cite{Goldenfeld1992}. For example, the equilibrium
liquid-gas transition belongs to the Ising universality class and has
a lower critical dimension of one.

\begin{figure*}[t]

\includegraphics{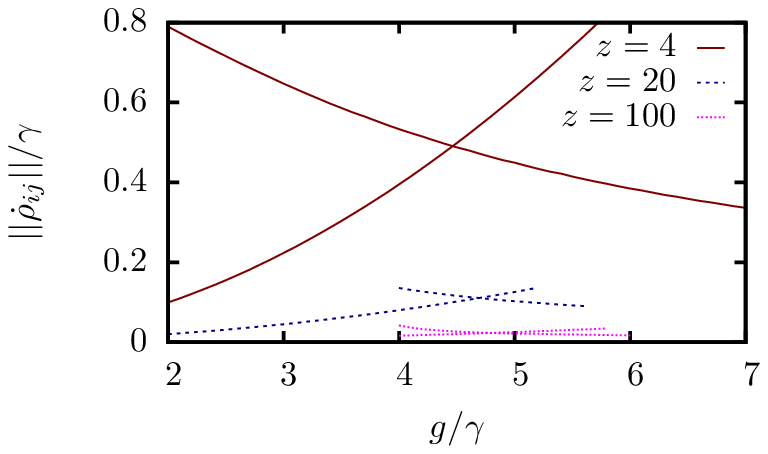} \includegraphics{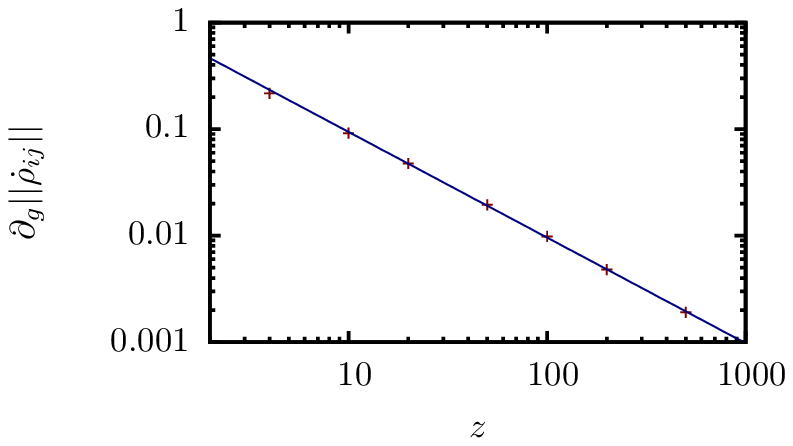}

\caption{Analysis of the dissipative Rydberg gas in the limit of large
  coordination number $z$. The left panel shows the position of the
  first-order transition indicated by the crossing of the two local
  minima in the variational norm. The right panel displays the slope
  of the variational norm right below the transition point. The solid
  line is an algebraic fit to the data with an exponent of
  $-0.99\pm 0.01$, perfectly consistent with a $1/z$ behavior.}

\label{fig:z}

\end{figure*}

In the following, we will investigate the dissipative Rydberg gas in
the limit where the coordination number $z$ goes to infinity. In this
case both the variational method and the first-order hierarchy
equations become exact, as the true steady state of the master
equation will be given by a product state. In particular, it is
instructive to look at this limit to investigate the role of the
bistability found in the solution of the hierarchy equations. For
this, we analyze the residual dissipation between the two local minima
found by the variational method. As shown in Fig.~\ref{fig:z}, the
variational method always yields a unique steady state. Here we find
that the larger the coordination number becomes, the smaller the
difference in residual dissipation. However, the regime of true
bistability indicated by a vanishing of the slope of the variational
norm is only reached asymptotically as $1/z$, and for any finite value
of $z$ (i.e., for finite spatial dimensionality), there is no
bistability. Consequently, it is incorrect to interpret the bistable
behavior predicted by the hierarchy equations as a signature of a
genuine thermodynamic phase, as it does not have a finite lower
critical dimension.

Rather, these findings suggest that the variational method is the
correct starting point from which arguments in favor of the existence
of thermodynamic phases in sufficiently high dimensions can be
based. This is of course in contrast with equilibrium systems, where
such arguments can be made based on a mean-field decoupling, which is
the equivalent of the first-order hierarchy equations.

It is worth mentioning that the situation is very different when the
hierarchy equations predict an antiferromagnetic phase. There, we also
find an antiferromagnetic phase within the variational method, albeit
with a smaller extension. While these results puts the existence of
such an ordered phase on firmer grounds, the role of quantum
fluctuation could still preclude its observation in actual
experiments, if the lower critical dimension of the transition is
three or larger.

\subsection{Superatom model}

We now extend the previous discussion of the lattice model of
Sec.~\ref{sec:lat} to the case of a continuum, as it has been the case
in the experimental situation in Ref.~\cite{Malossi2014}. In such a
case, the Rydberg blockade will ensure that Rydberg excitations
spontaneously form ordered structures
\cite{Weimer2008a,Schauss2012}. Although these correlations are
short-ranged, we may still well replace the underlying continuum by a
lattice structure with a lattice spacing corresponding to the typical
spacing between Rydberg excitations. We can then determine the lattice
spacing in a self-consistent manner \cite{Heidemann2008}, finding
\begin{equation}
  z\frac{C_6}{a^6} = \sqrt{\Omega_{\text{eff}}^2+(2\Delta)^2}.
\end{equation}
The factor of two in front of the detuning $\Delta$ can be understood
as realizing the antiblockade condition $C_6/a^6 = 2\Delta$ in the
limit of vanishing Rabi frequency. The effective Rabi frequency
$\Omega_\text{eff}$ is derived from a renormalization of the atomic
Rabi frequency $\Omega$ due to a (limited) collective enhancement. Far
away from resonance, the transition to the first Rydberg excitation is
still collectively enhanced, but the second Rydberg excitation can
then only appear at specific positions that satisfy the antiblockade
condition. Assuming there is always exactly one distance where the
antiblockade condition is fulfilled, we find that we can describe the
dynamics of such a superatom in terms of the number of atoms inside
the superatom, $N_s$, by renormalizing the Rabi frequency as
$\Omega_\text{eff} = N_s^{1/4}\Omega$, i.e., the geometric mean of the
Rabi frequency for the first and the second Rydberg excitation. 

In the following, we assume the underlying lattice to be a cubic
lattice (i.e, $z=6$), however, we would like to stress that our
results are basically independent of $z$, as long as it chosen
consistently with the assumed lattice structure.  Then, we can compute
the number of atoms per superatom to be
\begin{equation}
  N_s = n \sqrt{\frac{C_6}{2\sqrt{\Delta^2+\Omega^2}}},
\end{equation}
with $n$ being the density of ground state atoms. In the experimental
situation of Ref.~\cite{Malossi2014}, the system was either on
resonance or far away from it, i.e., either the condition $\Omega \gg
\Delta$ or $\Omega \ll \Delta$ has been fulfilled. Additionally, we
capture the experimental situation by including a dephasing term
associated with the laser linewidth $\kappa$, according to the jump
operators $c_i' = \sqrt{2\kappa} P_r^{(i)}$
\cite{Zoller1978}.

\begin{figure}[b]
\begin{tabular}{cc}
  \includegraphics[width=4.15cm]{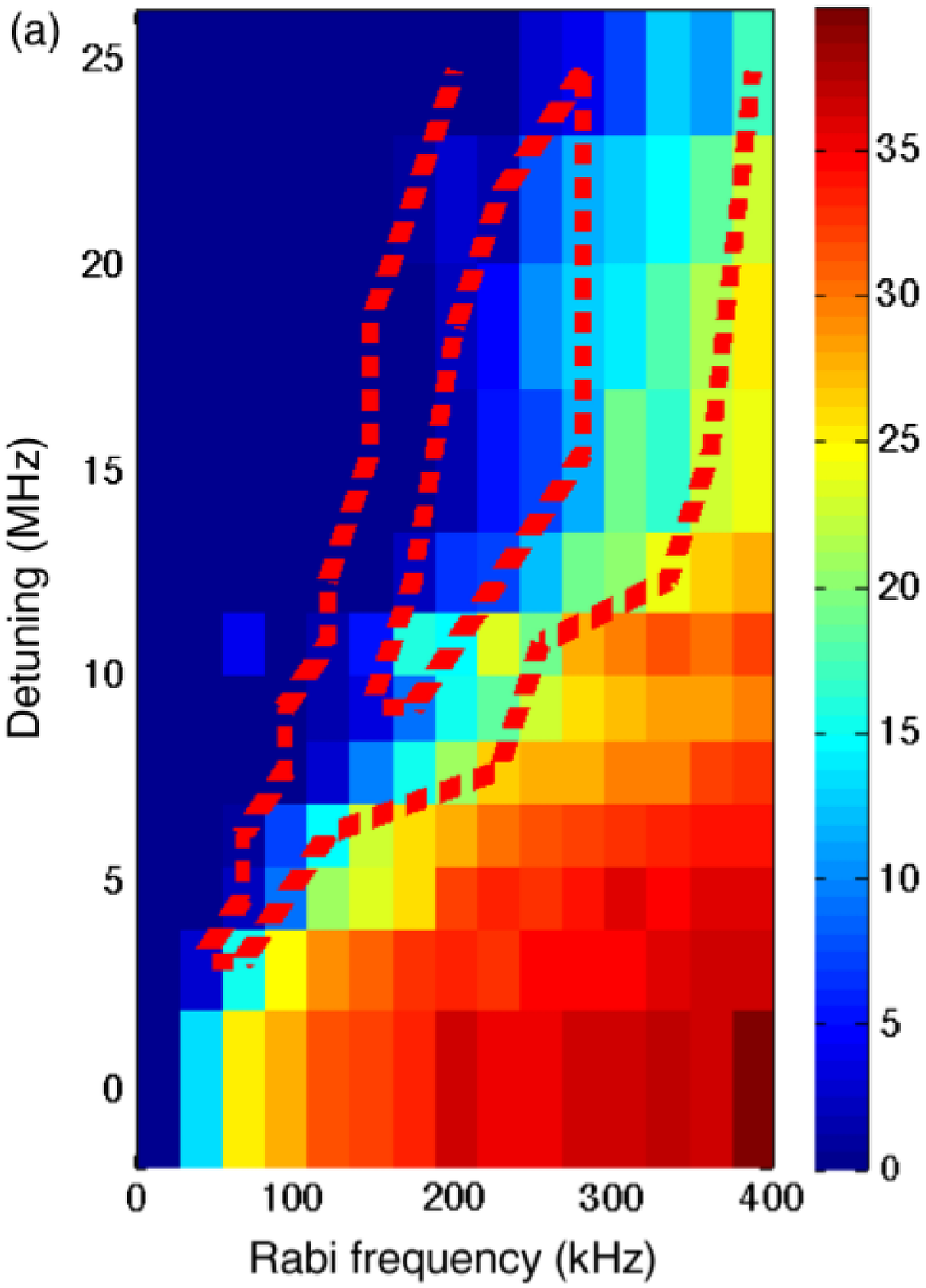} &
  \includegraphics[width=4.15cm]{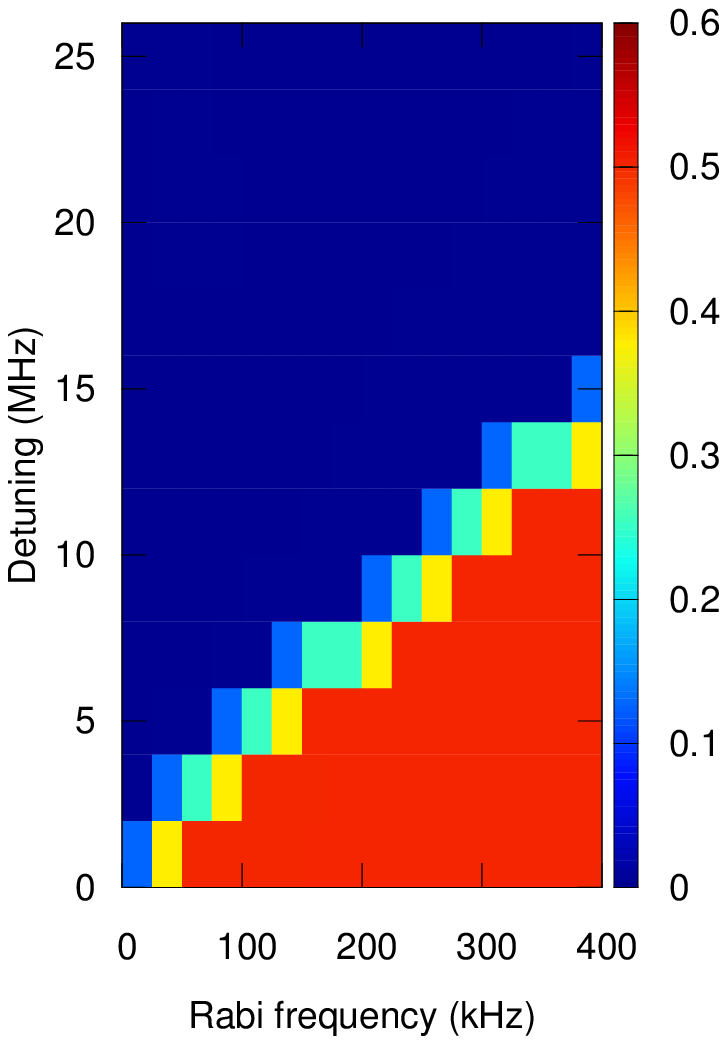}
\end{tabular}
\caption{Comparison of the stationary state of a dissipative Rydberg
  gas obtained by the Pisa experiment (left, taken with permission
  from \cite{Malossi2014}) and by the variational method (right). For
  the experimental data, the color coding refers to the total number
  of Rydberg excitations in the sample, while for the numerical
  simulations, it represents the fraction of excited superatoms
  ($\gamma = 5\,\mathrm{kHz}$, $\kappa = 500\,\mathrm{kHz}$, $n=
  1.8\times 10^{17}\,\mathrm{cm}^{-3}$, $C_6 = 7.54\times 10^{-58}
  \mathrm{Jm}^{-6}$).}
  \label{fig:pisa}
%  \vspace{-.5\baselineskip}
\end{figure}

We are now in the position to compute the steady state of the system
using the variational approach for correlated states. Remarkably, our
results are in good quantitative agreement with the experimental
observations in Ref.~\cite{Malossi2014}, see Fig.~\ref{fig:pisa}. The
jump in the density of Rydberg excitations corresponds to the
first-order phase transition between a low-density gas and a
high-density liquid, see Sec.~\ref{sec:lat}. In close analogy to the
classical liquid-gas transitions, the first-order transition ends in a
critical point. However, due to experimental limitations on the
dephasing rate and on the Rabi frequency, it appears that the
observation of the critical region of the dissipative Rydberg gas
remains a significant challenge.

\section{Summary}

In summary, we have given are detailed discussion of the recently
introduced variational principle for steady states of dissipative
quantum many-body systems \cite{Weimer2015}. We have exemplified its
usefulness by focusing on the driven-dissipative Rydberg gas, and we
have made a comparison of the variational approach to hierarchy
equation methods, finding severe conceptual advantages in favor of the
variational approach. Finally, we have found remarkable quantitative
agreement with experimental data for the phase transition between a
low-density gas of Rydberg excitations and a high-density liquid. Our
results strengthen the position of the variational method as a key
tool to analyze dissipative quantum many-body systems.

\begin{acknowledgments}

  We acknowledge fruitful discussions with O.~Morsch, T.~Osborne, and
  M.~Piani. This work was funded by the Volkswagen Foundation.

\end{acknowledgments}

%\bibliographystyle{aip}
%\bibliography{/home/hendrik/da/hendrik}
%\bibliography{/home/itp/weimer/hendrik}

%\clearpage

\end{document}